\documentclass[11pt]{amsart}
\usepackage{geometry}                
\usepackage{graphicx}
\usepackage{amssymb}
\usepackage{epstopdf}
\usepackage{fancyhdr}
\usepackage{hyperref}
\hypersetup{
	colorlinks=true,
	citecolor=blue
}
\usepackage[affil-it]{authblk}
\makeatletter
\def\@maketitle{%
  \newpage
  \null
  \vskip 2em%
  \begin{center}%
  \let \footnote \thanks
    {\Large\bfseries \@title \par}%
    \vskip 1.5em%
    {\normalsize
      \lineskip .5em%
      \begin{tabular}[t]{c}%
        \@author
      \end{tabular}\par}%
    \vskip 1em%
    {\normalsize \@date}%
  \end{center}%
  \par
  \vskip 1.5em}
\makeatother

\title{\sc{A Practioner's Guide to \\Evaluating Entity Resolution Results}}
\author{Matt Barnes \thanks{mbarnes1@cs.cmu.edu}}
  \affil{mbarnes1@cs.cmu.edu \\ School of Computer Science \\ Carnegie Mellon University}
\date{October, 2014}                                           

\begin{document}
\maketitle
\section{Introduction}
Entity resolution (ER) is the task of identifying records belonging to the same entity (e.g. individual, group) across one or multiple databases. Ironically, it has multiple names: deduplication and record linkage, among others. In this paper we survey metrics used to evaluate ER results in order to iteratively improve performance and guarantee sufficient quality prior to deployment. Some of these metrics are borrowed from multi-class classification and clustering domains, though some key differences exist differentiating entity resolution from general clustering. Menestrina et al. empirically showed rankings from these metrics often conflict with each other, thus our primary motivation for studying them \cite{Menestrina2010}. This paper provides practitioners the basic knowledge to begin evaluating their entity resolution results.

\section{Problem Statement}
Our notation follows that of \cite{Menestrina2010}. Consider an input set of records $I = \{a, b, c, d, e\}$ where $a, b, c, d,$ and $e$ are unique records. Let $R = \{\langle a, b, d\rangle, \langle c, e\rangle \}$ denote an entity resolution clustering output, where $\langle ...\rangle$ denotes a cluster. Let $S$ be the true clustering, referred to as the ``gold standard.'' The goal of any entity resolution metric is to measure error (or similarity) of $R$ compared to the gold standard $S$.

\section{Pairwise Metrics}
Pairwise metrics consider every pair of records as samples for evaluating performance. Let $Pairs(R)$ denote all the intra-cluster pairs in the clustering $R$. In our example, $Pairs(R) = \{(a, b), (a, d), (b, d), (c, e)\}$. Confusingly, some studies treat pairs only as those where a direct match was made and not matches made through transitive relations \cite{Michelson2009}. For example, \cite{Michelson2009} would exclude $(a,d)$ if the matches leading to $R$ were $a\approx b$, $b \approx d$, and $c \approx e$, where $\approx$ denotes a match. We choose the former definition because it is independent of the underlying matching process -- it only depends on the final entity resolution results.

Unlike many machine learning classification tasks, we never consider non-matches (i.e. inter-cluster pairs) in entity resolution metrics \cite{Christen2007}. In conventional clustering tasks, the number of clusters is constant or sub-linear with respect to the number of records $n$ \cite{Getoor2012}. However, the number of clusters is $\mathcal{O}(n)$ in conventional ER tasks. So though the number of intra-cluster pairs is $\mathcal{O}(n)$ (e.g. true positives), the number of inter-cluster pairs (e.g. true negatives) is $\mathcal{O}(n^2)$. To illustrate, consider our original example with 5 records and 2 clusters. There are 4 intra-cluster pairs and 6 inter-cluster pairs. Now, compare this to a larger database with 50 records and 20 clusters, all of comparable size to the original example. There will be approximately 40 intra-cluster pairs but likely over 2000 inter-cluster pairs. Thus, metrics using inter-cluster pairs (e.g. False Positive Rate) will improve exponentially with respect to the number of records in the database and provide overly optimistic results for large databases.

\subsection{Pairwise Precision, Recall, and F$_1$}
Using $Pairs$ as the samples, the pairwise precision and recall metric functions follow conventional machine learning definitions. The harmonic mean of these metrics leads to the most frequently used entity resolution metric, pairwise $F_1$. All these metrics are bound from $[0, 1]$.
\begin{equation}
PairPrecision(R,S) = \frac{|Pairs(R) \cap Pairs(S)|}{|Pairs(R)|}
\end{equation}
\begin{equation}
PairRecall(R,S) = \frac{|Pairs(R) \cap Pairs(S)|}{|Pairs(S)|}
\end{equation}

\begin{equation}
PairF_1(R,S) = \frac{2*PairPrecision(R,S)*PairRecall(R,S)}{PairPrecision(R,S) + PairRecall(R,S)}
\end{equation}

The benefit of pairwise metrics is their intuitive interpretation. Pairwise precision is the percentage of matches in the predicted clustering that are correct. Pairwise recall is the percentage of matches in the true clustering that are also in the predicted clustering. Unfortunately pairwise metrics may convey overly optimistic results, depending on the use case. For example, in many entity resolution tasks the end user only cares about the final entity -- not the records it comprises. Mismatching two singleton entities has an insignificant impact on pairwise metrics compared to incorrectly joining or splitting two large clusters. 

\vspace{\baselineskip}
\section{Cluster Metrics}
Like the pairwise metrics, all the cluster metrics discussed here are bound by $[0, 1]$, a convenient property when comparing across datasets and for setting quality standards.

\subsection{Cluster Precision, Recall, and $F_1$}
Cluster level metrics attempt to capture a more holistic understanding of the final entities. At the extreme opposite of pairwise metrics, cluster level precision \cite{Huang2006} and recall \cite{Wellner2004} consider \emph{exact} cluster matches. Mathematically, cluster precision and recall are defined as $\frac{|R \cap S|}{|R|}$ and $\frac{|R \cap S|}{|S|}$, respectively. Now, mismatching two singleton entities will have the same impact as mismatching two larger clusters. Obviously, this metric has the opposite drawback -- even one corrupted match in a cluster will cause an entire cluster to mismatch due to the use of exact comparisons. Thus, this metric is rarely used in favor of its predecessor, closest cluster precision, recall, and $F_1$.

\subsection{Closest Cluster Precision, Recall, and $F_1$}
Closest cluster metrics correct for the previous cluster-level drawbacks by incorporating a notion of cluster similarity \cite{Benjelloun2009}. Using the Jaccard similarity coefficient $J(r,s) = \frac{|r \cap s|}{|r \cup s|}$ to capture cluster similarity, the precision and recall can be expressed as

\begin{equation}
ccPrecision(R,S) = \frac{\sum_{r\epsilon R}\max{_{s\epsilon S}(J(r,s))}}{|R|}
\end{equation}
\begin{equation}
ccRecall(R,S) = \frac{\sum_{s\epsilon S}\max{_{r\epsilon R}(J(s,r))}}{|S|}
\end{equation}
where $r$ and $s$ are clusters in $R$ and $S$, respectively. This metric, and many of the ones following, attempt to balance the tradeoffs of the pairwise and exact cluster metrics.

\subsection{Purity and K}
Cluster purity was first proposed in 1998 \cite{Solomonoff1998} and later extended to Average Cluster Purity (ACP) and Average Author Purity (AAP) (archaically referred to as Average Speaker Purity) \cite{Ajmera2002}. The ACP and AAP are defined as

\begin{equation}
ACP = \frac{1}{N}\displaystyle\sum\limits_{r \epsilon R}\sum\limits_{s \epsilon S} \frac{|r \cap s|^2}{|r|}
\end{equation}
\begin{equation}
AAP = \frac{1}{N}\displaystyle\sum\limits_{r \epsilon R}\sum\limits_{s \epsilon S} \frac{|r \cap s|^2}{|s|}
\end{equation}
Then the $K$ measure is defined as the geometric mean of these values, $K = \sqrt{AAP * ACP}$. In many applications only a single purity metric is evaluated, usually something comparable to ACP. For example, \cite{Manning2008} considers the dominant class in each cluster by defining purity as $p = \frac{1}{N}\sum_{r \epsilon R}\max{_{s \epsilon S}|r \cap s|}$. The use of this single metric is misleading and only shows one half of the precision/recall coin. As an extreme example, setting $|R| = N$ (i.e. each record in its own cluster) would achieve a perfect $p = 1.0$, yet is clearly far from ideal.
 
\subsection{Homogeneity, Completeness, and V-Measure}
Homogeneity and completeness are entropy based metrics, somewhat analogous to precision and recall, respectively \cite{Rosenberg2007}. A cluster in $R$ has perfect homogeneity if all records belong to the same cluster in $S$. Conversely, a cluster in $S$ has perfect completeness if all its records belong to the same cluster in $R$. Entropy $H$ and its conditional variation are defined as

\begin{equation}
H(S) = -\frac{1}{|S|}\displaystyle\sum\limits_{s \epsilon S}\displaystyle\sum\limits_{r \epsilon R}|r \cap s| \log\frac{\sum_{r \epsilon R}|r \cap s|}{|S|}
\end{equation}

\begin{equation}
H(S|R) = -\frac{1}{N}\displaystyle\sum\limits_{r \epsilon R}\displaystyle\sum\limits_{s \epsilon S}|r \cap s| \log\frac{|r \cap s|}{\sum_{s \epsilon S}|r \cap s|}
\end{equation}
where $N$ is the total number of records. Using these entropies, homogeneity and completeness are defined as:

\begin{equation}
Homogeneity(R,S) = \begin{cases} 1 & \mbox{if } H(S)=0 \\
1 - \frac{H(S|R)}{H(S)} & \mbox{else} \end{cases}
\end{equation}

\begin{equation}
Completeness(R,S) = \begin{cases} 1 & \mbox{if } H(R)=0 \\
1 - \frac{H(R|S)}{H(R)} & \mbox{else} \end{cases}
\end{equation}
V-Measure is defined analogously to the $F_1$ metric as the harmonic mean of homogeneity and completeness.

\begin{equation}
V_\beta = \frac{(1+\beta^2)*Homogeneity(R,S)*Completeness(R,S)}{\beta^2*Homogeneity(R,S) + Completeness(R,S)}
\end{equation}
where $\beta$ is a user defined parameter, usually set to $\beta=1$ as in the $F_1$ metric. Completeness is weighed more importantly if $\beta>1$ and homogeneity is weighed more importantly if $\beta<1$. Some sources use $\beta$ instead of $\beta^2$ weighting, we chose the latter due to popularity.

\subsection{Other Metrics}
The natural language processing community uses several other entity resolution metrics, which are rarely using in machine learning and database applications \cite{Maidasani2012}. We refer the reader to MUC-6 \cite{Vilain1995}, $B^3F_1$ \cite{Bagga1998}, and CEAF \cite{Luo2005}.

\vspace{\baselineskip}
\section{Edit Distance Metrics}
Edit distance metrics can be thought of similarly to string edit distance functions. They are a measure of the information lost and gained while modifying $R$ to $S$. Unfortunately, they do not have the convenient $[0, 1]$ bound and are thus difficult to relate to any notion of a `good' score.

\subsection{Variation of Information}
VI \cite{Meila2003} can conveniently be expressed with the previous conditional entropy metric \cite{Rosenberg2007}.
\begin{equation}
VI(R,S) = H(S|R) + H(R|S)
\end{equation}
An important property of VI is it does not directly depend on  $N$, only the sizes of the clusters. Thus, it is acceptable to add records from new clusters to a database while continuously measuring VI performance.

\subsection{Generalized Merge Distance}
Generalized Merge Distance (GMD) is perhaps the most comprehensive metric in the sense it can be used to directly calculate several other metrics \cite{Menestrina2010}. $GMD(R,S)$ is the minimum legal path cost of converting $R$ to $S$, where the cost of splitting and merging sets of records are user-defined operation-order-independent functions. Many such functions exist, such as $f(x, y) = k$, $f(x, y) = kxy$, $f(x,y) = k_1 + k_2xy$ where $x$ and $y$ are the size of the record sets to split or merge and $k$ is a constant. We refer the reader to \cite{Hosszu1971} for a background of operation-order-independence functions.

Menestrina et al. not only show $GMD(R,S)$ can be computed in linear time, but explicitly show how pairwise precision, recall, $F_1$, and VI can be computed using specific cost functions. Depending on the choice of cost functions, GMD is likely dependent on $N$ (the cost functions used in the VI formulation are one exception) and difficult to compare across datasets of different sizes.

\subsection{Conclusion}
Simple examples show a promising pairwise metric may have poor cluster-level performance \cite{Michelson2009}. More rigorous analysis shows this is not only possible, but common across a range of applications \cite{Menestrina2010}. At an absolute minimum, we recommend evaluating with pairwise $F_1$ because of its simplicity and popularity. We also recommend the use of a cluster metric and Generalized Merge Distance -- which could conveniently be configured to calculate VI and the pairwise $F_1$ in linear time.

All the metrics discussed herein rely on the availability of a ``gold standard'' $S$. In practice, human-labeled results rarely number beyond several thousand samples. On large datasets, a relative gold standard may be obtained by foregoing blocking efficiency and running an exhaustive ER algorithm on the entire database \cite{Menestrina2010}. We note, however, that doing so on databases larger than even 10,000 records is infeasible for some algorithms \cite{Benjelloun2009}. Further, an exhaustive approach is still only an approximation and carries no guarantees relative to the true clustering. A need exists for semi- and un-supervised evaluation metrics. Some metrics exist for a very specific subset of circumstances, but for the majority of applications the general research problem is still open \cite{Winkler2006}.

\bibliography{metrics}{}
\bibliographystyle{ieeetr} 

\end{document}